\documentclass[acmlarge,prodmode,acmtosem]{acmsmall} 

\usepackage[ruled]{algorithm2e}

\usepackage{verbatim}
\usepackage{times}
\usepackage{multirow}
\SetAlFnt{\small}
\SetAlCapFnt{\small}
\SetAlCapNameFnt{\small}
\SetAlCapHSkip{0pt}
\IncMargin{-\parindent}
\usepackage{tabularx}

\acmVolume{X}
\acmNumber{X}
\acmArticle{X}
\acmYear{2019}
\acmMonth{1}

\usepackage{color} 
\usepackage{graphicx}
\usepackage{booktabs}
\usepackage{subcaption}
\usepackage{times}
\usepackage{makecell}
\usepackage{appendix}
\usepackage{collectbox}
\usepackage{mdframed}

\makeatletter

\usepackage[usenames,dvipsnames, table]{xcolor}

\usepackage{listings} 
\lstdefinelanguage{Uppaal}{ 
	basicstyle=\scriptsize\sffamily, 
	keywords={after_update,assign,before_update,break,case,const,continue,
			 default,else,enum,for,guard,if,meta,process,progress,return,select,
			 state,sync,switch,trans,system,while},
	keywords={[2]broadcast,bool,clock,chan,commit,init,int,scalar,struct,
			 typedef,urgent,void}, keywordstyle={[2]\bfseries},
	keywords={[3]false,true}, otherkeywords={[3]->},
	morekeywords={[3]->}, keywordstyle={[3]\bfseries},
	comment=[l]{//}, morecomment=[s]{/*}{*/}, 
	commentstyle=\itshape, 
	tabsize=4, 
	captionpos=b, 
	escapechar=@ 
}
\lstdefinelanguage[GUI]{Uppaal}[]{Uppaal}{ 
	keywordstyle={[2]\color{black!50!green}}, 
	otherkeywords={->}, keywordstyle={[3]\color{magenta}},
	commentstyle={\color{black!50!red}\itshape}, 
	literate={{-->}{$-->$}3} 
}
\lstdefinelanguage[LIT]{Uppaal}[GUI]{Uppaal}{ 
	literate={{->}{{$\leadsto$} }2 {-->}{{$\longrightarrow$} }2
	{=}{{$\gets$ }}2 {==}{{$==$}}2 {:=}{{$\gets$ }}2 {<=}{{$\leq$ }}2
	{>=}{{$\geq$ }}2 {&&}{{$\land$}}2 {||}{{$\lor$}}2 {<>}{{$\Diamond$}}1
	{[]}{{$\Box$}}1 {forall}{{$\forall$}}1 {exists}{{$\exists$}}1}
}

\AtBeginDocument{\setlength{\tempdimen}{\textwidth}}

\definecolor{verylightgray}{rgb}{0.9,0.9,0.9}
\definecolor{mauve}{rgb}{0.58,0,0.82}
\definecolor{jslightgray}{rgb}{.9,.9,.9}
\definecolor{jsdarkgray}{rgb}{.1,.1,.1}
\definecolor{jspurple}{rgb}{0.65, 0.12, 0.82}

\definecolor{customlightgray}{gray}{0.85}
\usepackage{array}
\usepackage{verbatim}
\usepackage{times}
\extrafloats{1000}

\renewcommand{\arraystretch}{1.2}
\title{Report on A Formally-Founded Model-Based Approach to Engineer Self-Adaptive Systems}
\author{
DANNY WEYNS
\affil{Katholieke Universiteit Leuven, Belgium \& Linnaeus University, Sweden}
M. USMAN IFTIKHAR
\affil{Katholieke Universiteit Leuven, Belgium \& Linnaeus University, Sweden}
}
\begin{abstract}
Self-adaptive systems manage themselves to deal with uncertainties that can only be resolved during operation. A common approach to realize self-adaptation is by adding a feedback loop to the system that monitors the system and adapts it to realize a set of adaptation goals. ActivFORMS (Active FORmal Models for Self-adaptation) provides an end-to-end approach for engineering self-adaptive systems. ActivFORMS relies on feedback loops that consists of formally verified models that are directly deployed and executed at runtime to realize self-adaptation. At runtime, the approach relies on statistical verification techniques that allow efficient analysis of the possible options for adaptation. Further, ActivFORMS supports on-the-fly changes of adaptation goals and updates of the verified models to to meet the new goals. ActivFORMSi provides a tool-supported instance of ActivFORMS. The approach has been validates using an IoT application for building security monitoring. This report provides complementary material to the paper ``ActivFORMS: A Formally-Founded Model-Based Approach to Engineer Self-Adaptive Systems'' [Weyns and Iftikhar 2019]. 
\end{abstract}
\category{Software engineering}{Software creation and management}{Designing software}
\terms{Design, Verification}

\begin{document}
\begin{bottomstuff}

\end{bottomstuff}
\maketitle

\section{Introduction}

Modern software systems are expected to deal with uncertain operating conditions, such as dynamics in work load or changes in the availability of resources. A common approach to handle such uncertainties is~\textit{self-adaptation}~\cite{Oreizy1998,Kephart2003,Garlan2004,Kramer2007,Cheng2009,Lemos2013,Weyns2019,WeynsBook2020}. Self-adaptation is realized by adding a feedback loop to the system that monitors the system and its environment and adapts the system in order to realize particular quality requirements (i.e., adaptation goals). A typical example is an elastic Cloud platform that monitors the applications of clients and automatically adapts its capacity to maintain its performance at the lowest possible cost.
Our focus is on \textit{architecture-based adaptation} that provides an effective approach to mitigate uncertainties at runtime~\cite{Oreizy1998,Garlan2004,Kramer2007,Weyns2012-1,10.1016/j.jss.2015.09.021}. Central to architecture-based adaptation is a feedback loop that uses runtime models to reason about adaptation~\cite{Garlan2004,5280648,Weyns2012-1}. A well-known approach to structure the feedback loop is by means of four basic elements: Monitor, Analyze, Plan, and Execute that share Knowledge models, commonly known as MAPE-K~\cite{Kephart2003,Dobson2006,exp}.  

A key challenge in the engineering self-adaptive systems is providing guarantees that the system complies with its adaptation goals~\cite{Camara2013,Cheng2014,Lemos2017,WeynsBook2020}. Different  approaches have been presented to provide such guarantees~\cite{Weyns2012-2,Tamura2014,Cheng2014,Weyns:2016}. We focus here on a popular approach that relies on the use of formal modeling and verification techniques at runtimer~\cite{Cheng2009,Calinescu2011,Lemos2013,Lemos2017,DBLP:conf/sigsoft/MorenoCGS15}. 
State of the art approaches have a number of limitations that hamper their application in practical systems. First, existing approaches typically rely on exhaustive verification techniques for runtime analysis, which suffers from the state explosion problem~\cite{Clarke2008}. Second, most existing approaches do not consider the correctness of the feedback loop behavior itself. Third, approaches do provide limited or no support for online changes of adaptation goals, which is generally considered an important type of uncertainty~\cite{Kramer2007,Sawyer2010,Souza:2013,9196226}. 

To tackle these limitations, we propose ActivFORMS, short for Active FORmal Models for Self-adaptation. ActivFORMS provides a formally-founded model-driven approach for engineering self-adaptive systems that spans four main stages of the life cycle of a feedback loop. At design time, a feedback loop is specified using formal models that are verified against a set of correctness properties. At deployment time, the models of the feedback loop models are directly deployed for execution using a model execution engine. At runtime, the feedback loop selects adaptation options that realize the adaptation goals in an efficient manner using statistical model checking. Finally, at 
evolution time, ActivFORMS offers basic support for online changing adaptation goals and updating the feedback loop model to meet the new goals. ActivFORMSi provides a tool-supported instance of ActicFORMS. We have validated ActivFORMSi to an Internet of Things (IoT) application for security monitoring of an area that is deployed at the Computer Science campus of KU Leuven by VersaSense\footnote{www.versasense.com/}.

ActivFORMS supports anticipated uncertainties of the managed system, the environment, the feedback loop, and the adaptation goals~\cite{Malek2011,Perez-Palacin:2014:UMS:2568088.2568095,MAHDAVIHEZAVEHI201745,9196226}. Our focus here is on parametric uncertainties, i.e., uncertainties that can be expressed as parameters of runtime models. ActivFORMs leverages on initial work that focused on correct behavior of MAPE models~\cite{Iftikhar2014,Didac2015}, simulation at runtime to analyze adaptation options~\cite{7573167}, and trustworthy self-adaptive systems~\cite{Calinescu2017}. All material related to ActivFORMS is available at the ActivFORMWS website.\footnote{https://people.cs.kuleuven.be/danny.weyns/software/ActivFORMS/} 

This report provides additional material to~\cite{abs-1908-11179} that describes the ActivFORMS approach in depth. We start the report with a brief summary of ActivFORMS and its instantiation. Then we present the additional material, incl. a series of examples that illustrate different stages of ActivFORMS for a health assistance example, a series of examples that illustrate different stages of ActivFORMS for an IoT application, the definition of selected parts of the feedback loop model, evaluation results on the tradeoff between accuracy and adaptation time for the IoT application, and finally a summary of opportunities for future research. 

\section{The ActivFORMS Approach in a Nutshell}\label{section:overview}

ActivFORMS offers a reusable end-to-end approach to engineer self-adaptive software system that are based on MAPE feedback loops~\cite{Kephart2003,Dobson2006,Calinescu2011,exp}. Other types of feedback loops, for instance based on principles from control theory~\cite{Shevtsov17} are not supported by ActivFORMS.

\begin{figure}[t!]
    \centering
    \includegraphics[width=\textwidth]{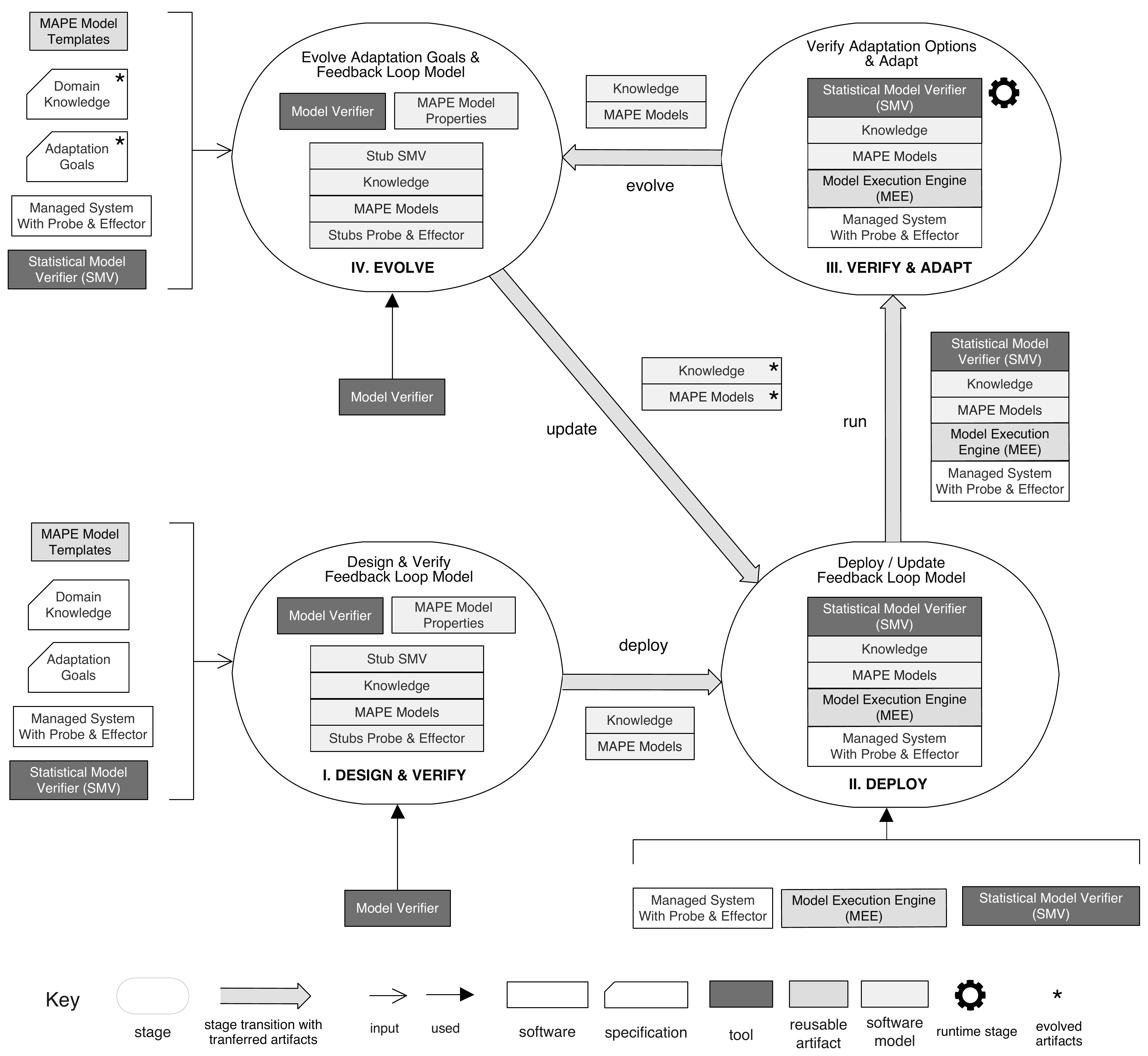}
    \caption{The four stages of ActivFORMS: I. Design \& Verify, II. Deploy, III. Verify \& Adapt, and IV. Evolve}
    \label{fig:activFORMS}
\end{figure}

Fig.~\ref{fig:activFORMS} gives a high-level overview of ActivFORMS.  The stages Design \& Deployment cover the design and enactment of a feedback loop leveraging MAPE model templates and a model execution engine. The Runtime stage realizes the adaptation of the managed system during operation to achieve the adaptation goals. To that end, ActivFORMS relies on statistical model verification. The Finally, the evolution stage realizes the evolution of feedback loops to deal with new or changing adaptation goals and updating runtime models. 

\textit{ActivFORMSi} offers a concrete instance of ActivFORMS.  Table~\ref{instantation-summary} summarizes the instantation. 

\begin{table}[h!]\caption{Summary of instantiation of  ActivFORMS with ActivFORMSi}\label{instantation-summary}
	\centering
	\begin{scriptsize}
		\renewcommand{\arraystretch}{1.2}
		\setlength{\tabcolsep}{0.4em}
		\begin{tabular}{p{3cm}p{9.5cm}}
			\Xhline{0.6pt}
			\textbf{ActivFORMS}   
			&  \textbf{ActivFORMSi}
			\\ \Xhline{0.6pt}
			MAPE model templates 
			& MAPE model templates based on timed automata with Uppaal  suite~\cite{Behrmann2004} \\ 	\Xhline{0.4pt}
			Quality models 
			& Stochastic timed automata  
			\\ 	\Xhline{0.4pt}
			Model execution engine 
			& Trusted virtual machine to execute timed automata 
			\\ 	\Xhline{0.4pt}
			Runtime analysis
			& Statistical model checking using Uppaal-SMC~\cite{David2015}
			\\ 	\Xhline{0.4pt}
			Goal management 
			& Trusted online update manager 		`			%
			\\ \Xhline{0.6pt}
		\end{tabular}
	\end{scriptsize}
\end{table}

\section{Example Applications}\label{section:examples}

We briefly introduce two example cases that we use to illustrate the different stages of ActivFORMS in the next sections: a health assistance system~\cite{Weyns2015c} and an IoT system~\cite{Iftikhar2017}.  

\subsection{Health Assistance System}
\begin{figure}[h!]
 	\centering	
 	\includegraphics[width=.75\linewidth]{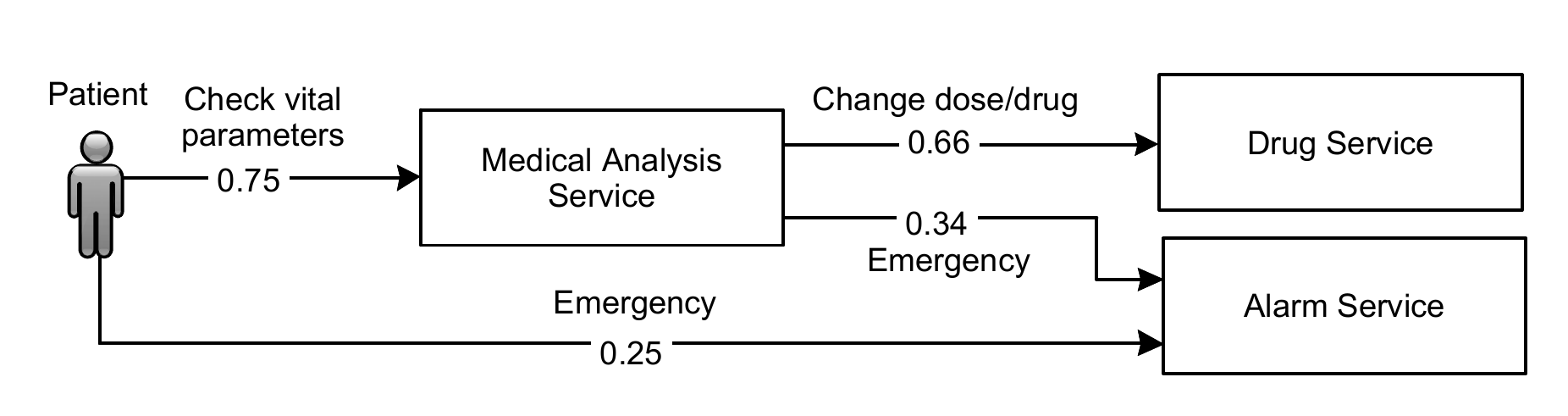}
 	\caption{Workflow of service-based health assistance system}
 	\label{fig:tas}
 \end{figure}

We consider a simple service-based health assistance system as shown in~Fig.~\ref{fig:tas} that is based on the TAS exemplar~\cite{Weyns2015c}. A Medical Service receives messages from patients with values of vital parameters. The service analyses the data and either instructs a Drug Service to notify a local pharmacy to deliver new medication to the patient or change the dose of medication, or it instructs an Alarm Service in case of an emergency to visit the patient by medical staff. The Alarm Service can also directly be invoked by a user via a panic button. The numbers associated with arrows in the workflow represent probabilities that actions are invoked. These probabilities represent uncertainties that may change over time. Each service can be implemented by a number of providers that offer services with different reliability (service failures), performance (response time), and cost (to use a service). The different properties of services may change at runtime, for example due to changing workloads at the provider side or unexpected failures of the communication network. Hence, these properties represent another type of uncertainty. At runtime, it is possible to pick any of the services offered by the providers. The aim of adaptation is to select dynamically services such that the average failure rate remains below a given threshold, while the cost is minimized.
 
\subsection{IoT System}
 
\begin{figure}[h!tb]
 	\centering
 	\includegraphics[width=0.9\textwidth]{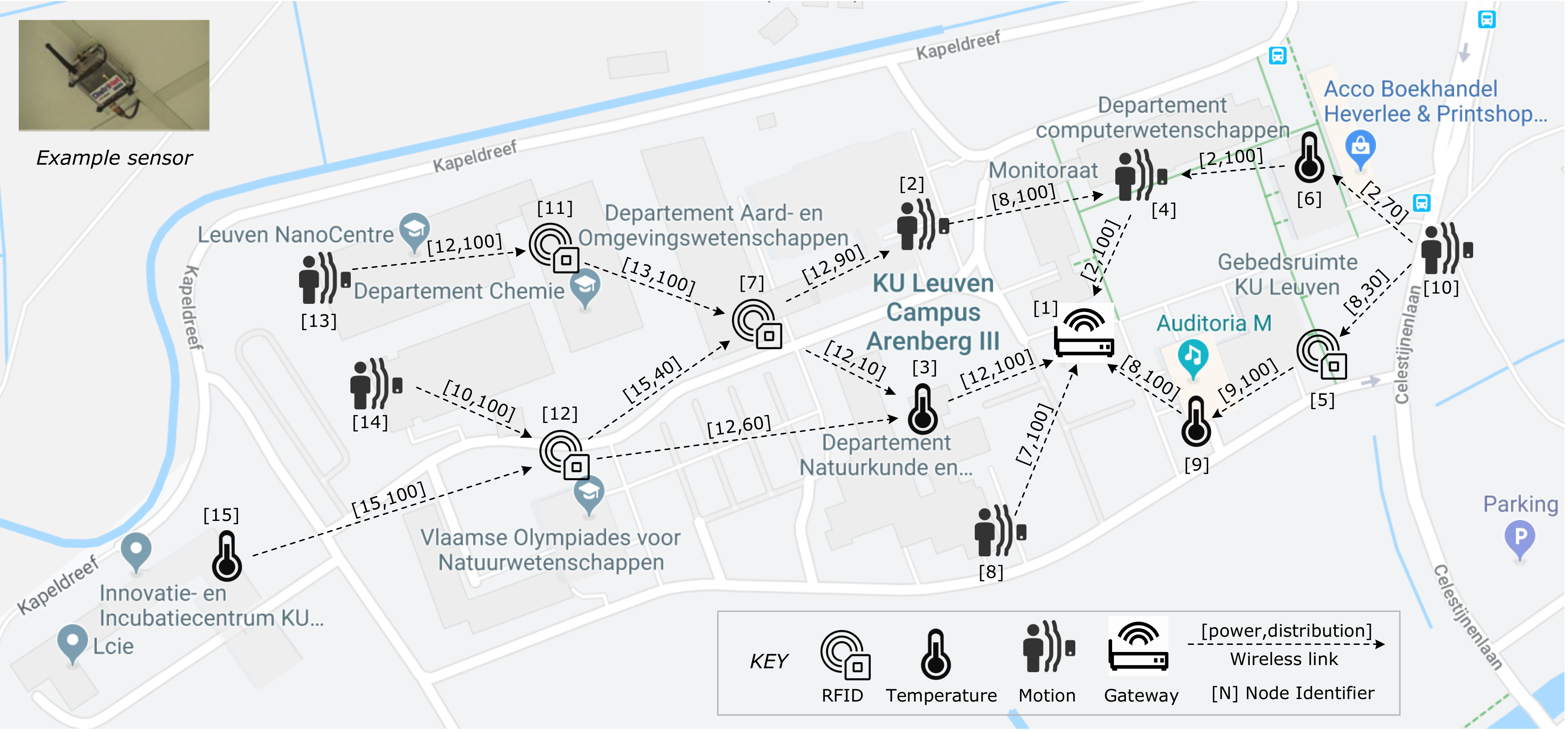}
 	\caption{DeltaIoT system with network topology and example of a sensor}
 	\label{fig:DeltaIoT}
 \end{figure}

We consider an IoT application, called DeltaIoT~\cite{Iftikhar2017}, see Fig.~\ref{fig:DeltaIoT}. The application consists of a collection of 15 battery-powered LoRa-based\footnote{https://www.lora-alliance.org/What-Is-LoRa/Technology} motes, each equipped with a sensor that senses a property in the environment, and facilities for wireless communication to send the data to a gateway that is deployed at a central monitoring facility. At that facility, campus staff can monitor the status of buildings and labs and take action whenever unusual behavior is detected.
 
 DeltaIoT uses multi-hop communication, where the communication is organized in cycles. Each cycle (e.g., of 8 minutes) consists of a number of communication slots (e.g., 40 slots), each slot enabling a sender mote and a receiver mote to communicate. 
 The IoT application is subject to noise caused by external factors such as weather conditions and fluctuating traffic load (e.g., packets produced by a passive infrared sensor are based on the detection of motion of humans). These uncertainties are difficult to predict upfront. 
The quality requirements for the network are $R1$: The average packet loss per period of 12 hours should not exceed 10\%, and $R2$: The energy consumption should be minimized. An additional adaptation goal  should keep the average latency of packets per 12 hours below 5\% of the cycle time ($R3$). This goal should be added to the system during operation. Finally, if no valid adaptation option is available, a reference setting should be applied; i.e., the transmission power of all motes should be set to maximum and all packets should be send to all parents for each mote ($R4$). 

The gateway provides an interface to monitor the network for each  cycle, including the traffic generated by a mote (number of messages sent from 0 to 10), the  energy consumed (in Coulomb), the settings of the transmission power that a mote used to communicate with each of its parent (in a range from 0 to 15), the distribution factor per mote and being the percentage of the packets sent by a source mote over the link to each of its parents (0 to 100\%), the packet loss (fraction of packets lost in the network [0...1]), the energy consumed by the network (Coulomb), and latency of the network (the fraction of the cycle time as a percentage that packets remain in the network). The interface can also be used to set the parameters of the network, i.e., the transmission power to be used by the motes to communicate via each link (0 to 15), and the distribution factor for motes with two parent (0 to 100\% in steps of 20\%). Finally, the network settings can be set to predefined values (i.e., the reference setting) that can be used as failsafe fallback. 

\section{ActivFORMS Aplied to Health Assistance System}

\subsection{Stage I: Design and Verify Feedback Loop Model}

In the first stage of ActivFORMS, a formally verified feedback loop model of the self-adaptive system is developed that includes a specification of \textit{Knowledge} and \textit{MAPE Models}, see Fig.~\ref{fig:activFORMS}.

\subsubsection{Design Feedback Loop Model}

To design a feedback loop model for the health assistance system, the designer requires different types of domain knowledge, such as the sample rate of vital parameters, usage patterns of the panic button, a list of initially available services with their characteristics. The designer also needs to understand the workflow of the service-based system and how this workflow can be monitored and adapted. Domain knowledge can be obtained in different ways, for instance by consulting with stakeholders, based on historical information, or through inspection of the code. One of the initial adaptation requirements defined by the stakeholders is to keep the average failure rate below a given value (in this example, we only consider this requirement). The adaptation goal for this requirement can be specified as a threshold of a parameter that represents the failure rate of the system. With this domain knowledge, the model of the workflow, and the specification of the adaptation goal at hand the designer can specify a feedback loop.

\begin{figure}[h!tb]
	\centering
	\includegraphics[width=0.9\textwidth]{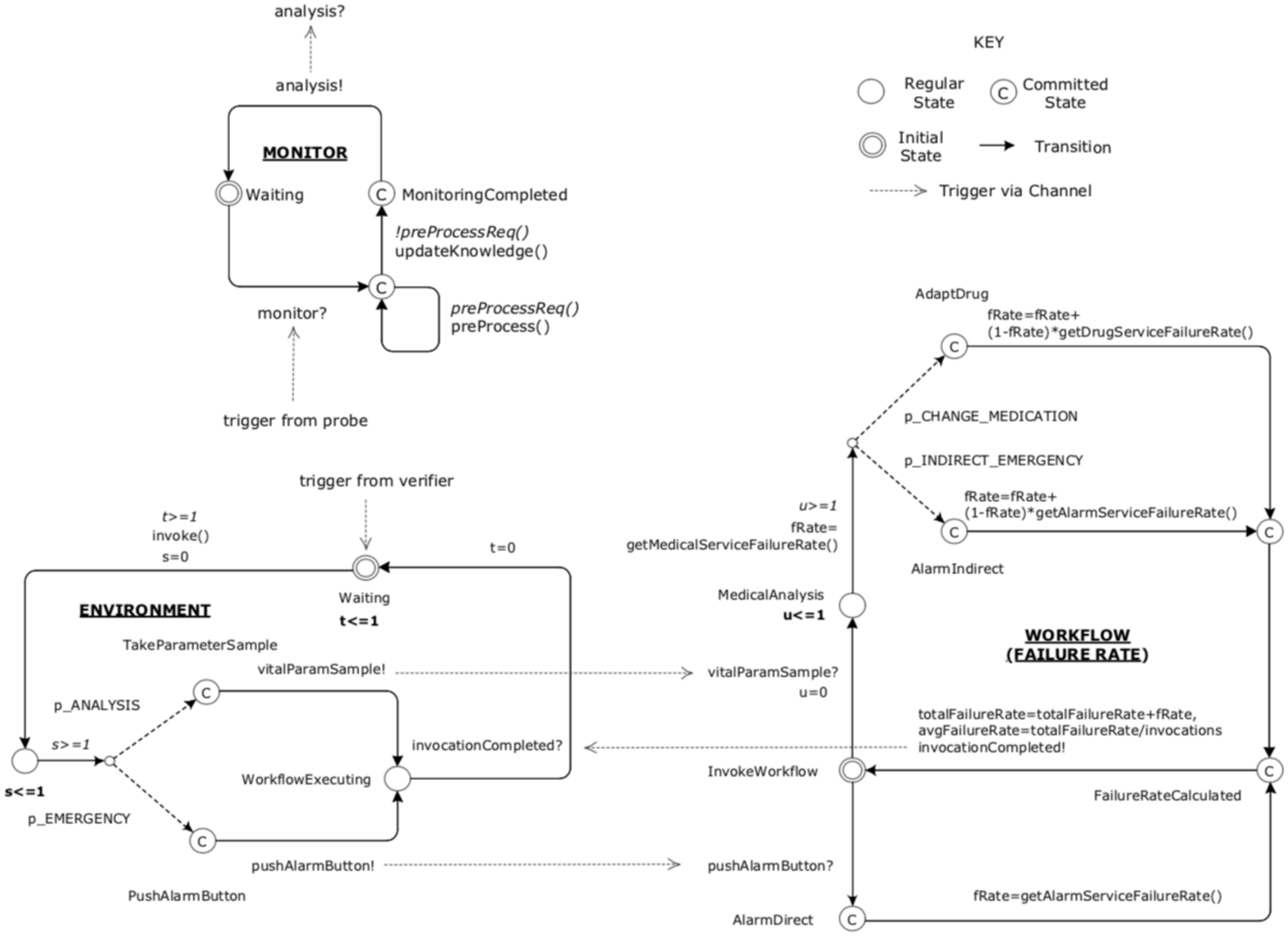}
	\caption{Example models of the feedback loop for the health assistance system}
	\label{fig:exampleMAPEmodels}
\end{figure}

Fig.\,\ref{fig:exampleMAPEmodels} shows a selection of models for the health assistance system that are specified as as a network of timed automata~\cite{ALUR1994183,David2015}.\footnote{Note that not all synchronization actions (“?” and “!” respectively) for all models are shown as only a subset of models are used in the paper. Yet, all models are available at the ActivFORMS website: https://people.cs.kuleuven.be/danny.weyns/software/ActivFORMS/
	For clarity, we write the guards in \textit{italic} font and the invariants in \textbf{bold} font in all automata models in this paper.
} The model of the monitor is part of the feedback loop. The models of the \textit{environment} and \textit{failure rate} are part of the knowledge; these models are used by the feedback loop at runtime to perform analysis of the adaptation options. The monitor model is based on model templates described in~\cite{Didac2015}. These templates have the same structure as the models shown in the figure, but the functions are abstractly defined and need to be instantiated for the problem at hand by the designer. As an example, the abstract function \textit{updateKnowledge}$()$ of the monitor is used to update relevant parameters of the knowledge. In the example, the designer needs to instantiate this function such that it updates the failure rate, the actual rate of panic button invocations, among other variables, whenever the monitor is triggered by the probe.

For the \textit{environment} model (which is used for online verification, see Stage III), the designer specifies the relevant external behavior, i.e., either a sample of the vital parameters is taken with a probability of \textit{p\_ANALYSIS} or the user pushes the alarm with a probability \textit{p\_EMERGENCY}. Depending on these actions the appropriate part of the quality model is triggered that models the workflow of the service system. For the \textit{failure rate} model, the designer specifies how failure rates are estimated. Depending on the action (determined in the environment model), either an alarm service is directly invoked (by the user) or a medical analysis service is invoked to analyze the vital parameters. In the former case, the failure rate is simply determined by the selected alarm service. In the latter case, the analysis results in either a change of the medication with probability  \textit{p\_CHANGE\_MEDICATION} or an alarm is activated with probability \textit{p\_INDIRECT\_EMERGENCE}. 
The failure rates are then determined based the combination of invoked services. The failure rate model can be instantiated for different service combinations enabling to predict their failure rate. The probabilities of the models of the environment and failure rate are kept up to date by the monitor based on information from the service providers.

\subsubsection{Test Knowledge Models}

\begin{figure}[h!tb]
	\centering
	\includegraphics[width=0.75\textwidth]{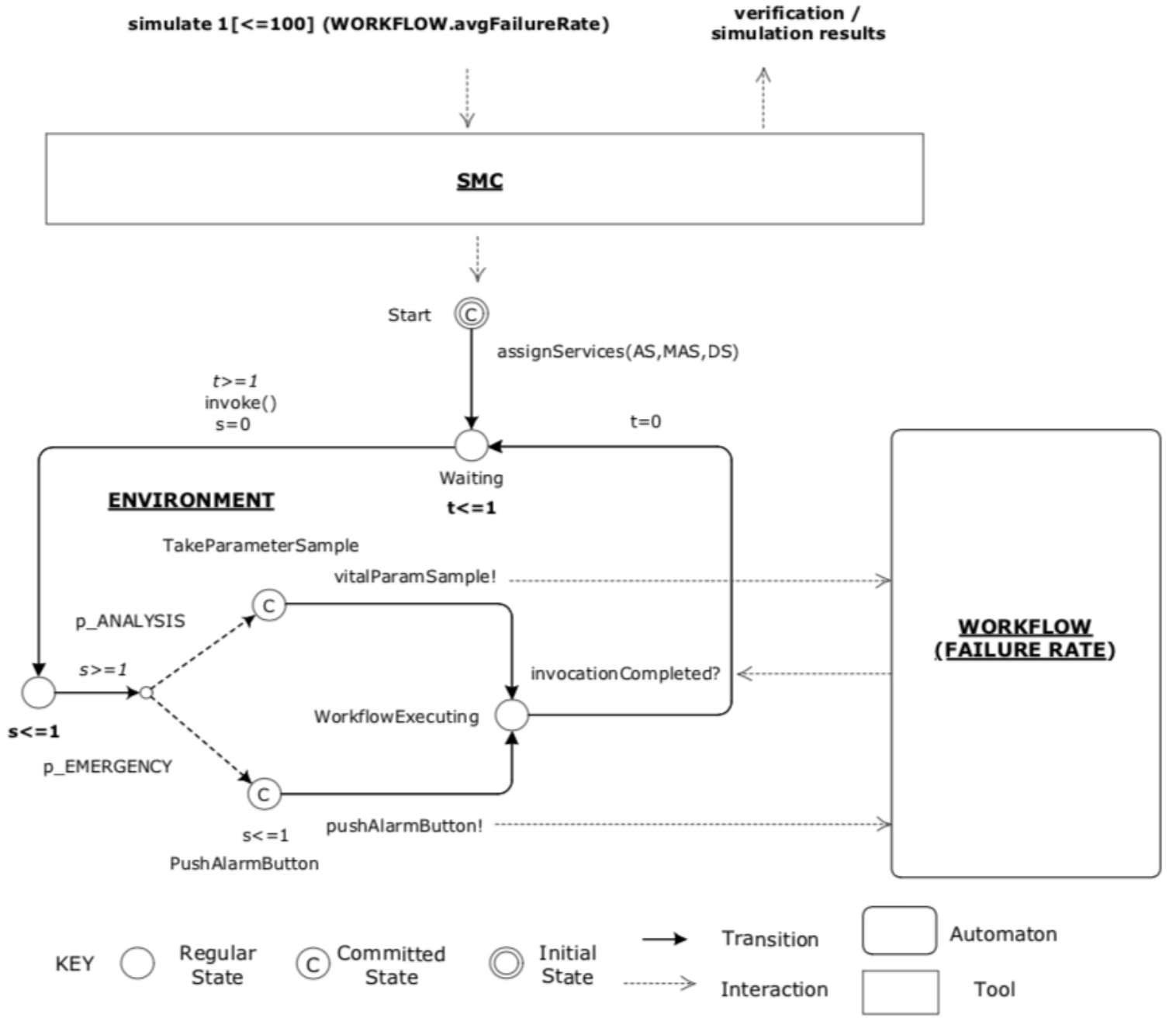}
	\caption{Example setup for testing quality models for the health assistance system}
	\label{fig:exampleTestMAPEmodels}
\end{figure}

\begin{figure}[h!tb]
	\centering
	\includegraphics[width=0.7\textwidth]{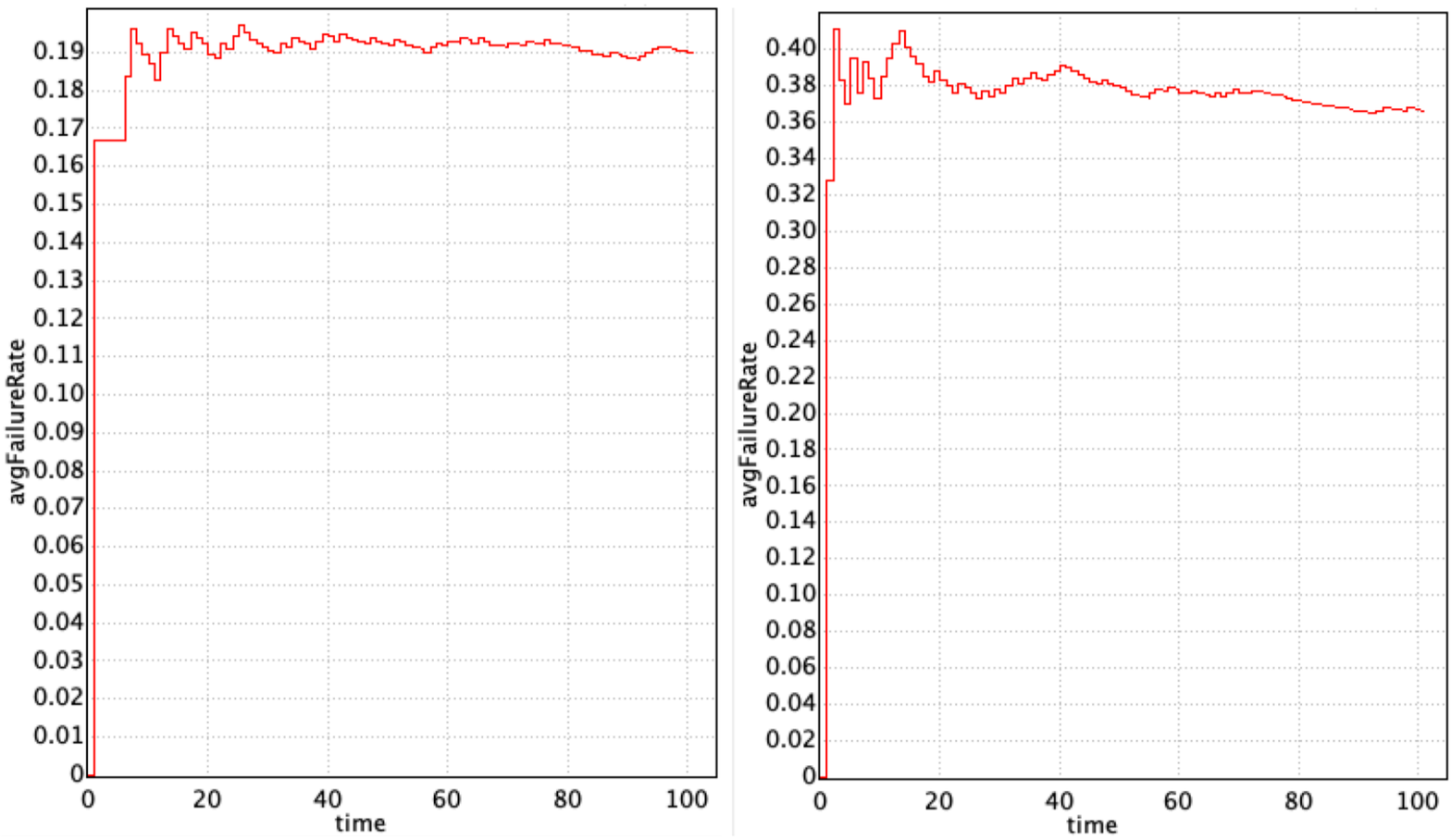}
	\caption{Excerpt of test results of the Knowledge models of the health assistance system. Left: results with regular values of failure rates.  Right: results with double values of failure rates.}
	\label{fig:resultsExampleTestMAPEmodels}
\end{figure}

Testing the knowledge models of the health assistance system requires basic information of the available services, user behavior and typical analysis results of vital sample analysis. Based on this information a test setup with representative input can be defined for the models of the environment and failure rate shown in Fig.~\ref{fig:exampleTestMAPEmodels}. The automata can then be tested using a tool, such as the Uppaal simulator~\cite{David2015}. Fig.~\ref{fig:resultsExampleTestMAPEmodels} illustrates this for two knowledge models of the health assistance system.

In this particular setting, we use a query that evaluates the average failure rate of the health assistance service over 100 runs. Listing~\ref{simulation} illustrates a number of settings that we used for the tests. The first line defines typical values for the probabilities of the models. Then follows two settings for failure rates of the different service instances, the first one with regular values of failure rates, the second one with double values.

\lstset{caption={Health assistance probe and effector methods.},label=simulation}
\footnotesize
{\ttfamily
	\begin{lstlisting}
//Probabilities of runtime parameters
const int p_EMERGENCY = 22, p_INDIRECT_EMERGENCY =34, 
p_ANALYSIS = 78, p_CHANGE_MEDICATION = 66;
...
//Failure rates of service instances
const double sFailureRates1[1,MAX_TYPE][1,MAX_IMP] =
  {{0.11,0.04,0.18,0.08},{0.12,0.07,0.18,0.10,0.15},{0.01,0.03,0.05,0.07,0.02}}
...
//Alternative setting for failure rates of service instances 
const double sFailureRates2[1,MAX_TYPE][1,MAX_IMP] =
  {{0.22,0.08,0.36,0.16},{0.24,0.14,0.36,0.20,0.30},{0.02,0.06,0.10,0.14,0.04}}
	\end{lstlisting}
}
\normalsize

\subsubsection{Verify MAPE Models}

\begin{figure}[h!tb]
	\hspace{-50pt}
	\includegraphics[width=1.2\textwidth]{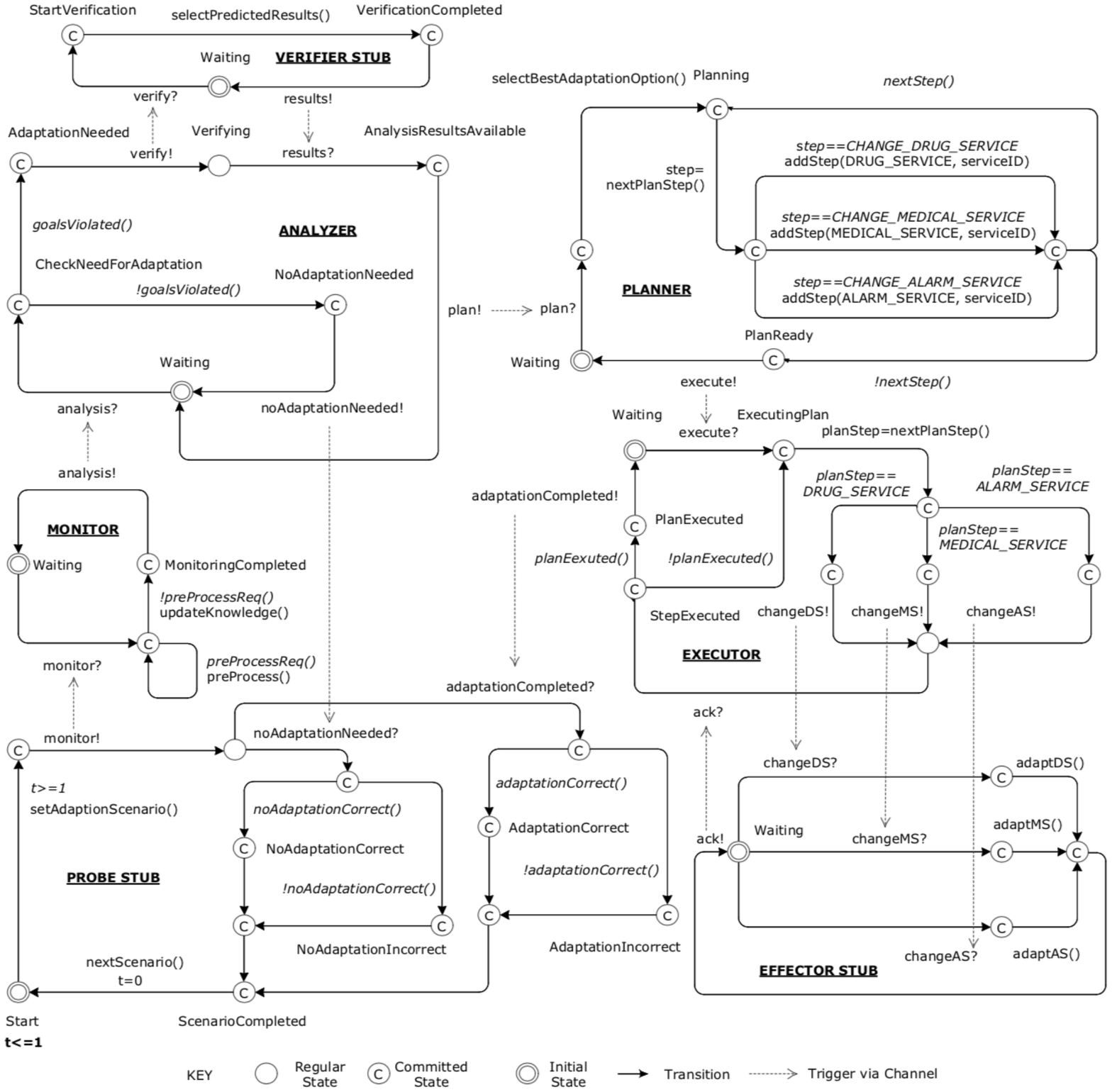}
	\caption{Verification of the MAPE models for the health assistance system}
	\label{fig:verification_setup}
\end{figure}

Fig.~\ref{fig:verification_setup} shows a set up for the verification of the MAPE models for the health assistance system, where the designer has specified stubs for the probe, effector, and verifier. The probe makes a distinction between scenarios that require adaptation and scenarios that do not require adaptation. For a scenario that requires adaptation, the parameter of the failure rate of the system configuration is set to a value that violates the goal. This will prompt the analyzer to predict the quality properties of the adaptation options using the verifier. The verifier stub will return a predefined set of analysis results for the running configuration, of which the best will be selected by the planner. The configuration is then adapted using the effector stub. Control is then returned to the probe stub that will check whether the adaptation is applied correctly, i.e., whether the adaptation option was selected with the optimal cost.

To check that no incorrect adaptation is applied for a set of scenarios, the designer can verify the property: 

\begin{quote}
	\textit{E $<>$ !Probe.AdaptationIncorrect}
\end{quote}

To check that the feedback loop correctly identifies the need for adaptation scenario when the adaptation goal for failure rate is violated for a given scenario, the following property can be verified:

\begin{quote}
	\textit{Monitor.MonitorCompleted\,$\,\,\&\&$\,\,Knowledge.failureRate $>$ Knowledge.fRateGoal\,\,$\rightarrow$\\
		\mbox{\ \ \ \ \ \ \ \ \ \ \ \ \ \ \ \ \ \ \ \ \ \ \ \ \ }	Analyzer.AdaptationNeeded}
\end{quote}

These properties can be verified using a model checking tool, such as Uppaal~\cite{David2015} or any other verification tool that supports the specification languages used for the models and properties.

For a scenario that requires no adaptation, the parameter of the failure rate of the system configuration is set to a value that satisfies the goal. A similar approach can then applied to check that the MAPE loop behaves correctly using the following property:

\begin{quote}
	\textit{Monitor.MonitorCompleted\,$\,\,\&\&$\,\,Knowledge.failureRate $<=$ Knowledge.fRateGoal\,\,$\rightarrow$\\
		\mbox{\ \ \ \ \ \ \ \ \ \ \ \ \ \ \ \ \ \ \ \ \ \ \ \ \ }	Analyzer.NoAdaptationNeeded}
\end{quote}

To obtain the required coverage of the tests, the designer can specify different sets of scenarios that do and do not require adaptation, ensuring that all the paths through the MAPE models are exercised.

\subsection{Stage II: Deploy and Enact Feedback Loop with Model Execution Engine}

In the second stage of ActivFORMS, the verified feedback loop model is deployed and enacted using a \textit{model execution engine}, see Fig.~\ref{fig:activFORMS}. 

\subsubsection{Deploy Feedback Loop Model}

For the feedback loop model of the health assistance system, a model execution engine is required that can execute a network of timed automata~\cite{Iftikhar2016}. The correctness of this engine relies on extensive testing.\footnote{We refer to the ActivFORMS website for details on the model execution engine.} When the engine loads a feedback loop model, it transforms the automata models into a graph representation that the engine can execute. The engine comes with template classes that the developer can use to realize the connections with external elements. These include classes to connect the MAPE models with probes and effectors, and a class to connect the analyzer model with the Uppaal-SMC tool \cite{David2015} that can be used to for the analysis of the quality models. 

\subsubsection{Enact the Model Execution Engine}

When the model execution engine and the feedback loop model are deployed and the connections are established (with the probe, effector, and verifier), the model execution can be started. We illustrate this activity for the DeltaIoT in the next section. 

\subsection{Stage III: Runtime Verification of Adaptation Goals and Decision Making}\label{runtime}

Stage III is a runtime stage where the verified feedback loop model executed by the execution engine monitors the managed system and its environment and adapts the managed system to realize the adaptation goals.

\subsubsection{Runtime Architecture of ActivFORMS} 

In the health assistance system, the managed system is the service infrastructure with the workflow that offers a probe to obtain data about the behavior of users, the actual quality properties of the system, the characteristics of different service instances, etc., and an effector to select concrete instances for the different services that are used by the workflow. The available service combinations determine the set of adaptation options. Change management comprises the feedback loop model with the model execution engine and statistical model verifier as we illustrated in the first two stages. We illustrate goal management in Stage IV below.

\subsubsection{Analysis of the Adaptation Options}\label{subsec:analysis}
 
Analysis consists of four steps. In step one, the adaptation options are composed by combining the different service instances for the workflow (the size of the adaptation space may be reduced by eliminating service instances that have shown poor qualities in the recent past). In step two, the values of the uncertainties are assigned based on recent data, i.e., the probabilities associated with different paths in the workflow. In step three, a statistical model checker verifies the quality models of the system, i.e., one model for failure rate and another for cost. Finally, in step four, the analyzer collects the estimates of the failure rate and cost of each adaptation option and updates the knowledge repository accordingly.

\subsubsection{Decision Making}

The health assistance system has two adaptation goals: a threshold goal for failure rate and an optimization goal for cost. A simple decision-making mechanism can apply the goals sequentially, i.e., first the adaptation options with a failure rate below the threshold are selected; next the adaptation option with the lowest cost of this subset is selected for adaptation. If none of the adaptation options complies with the failure rate goal, no adaptation may be applied, or alternatively a predefined set of services may be selected to adapt the system.

\subsection{Stage IV: Evolution of Adaptation Goals and Feedback Loop Model}\label{subsection:stage-4}

The fourth stage of ActivFORMS offers basic support for on-the-fly changes of the adaptation goals and the feedback loop model through the goal management layer~\cite{Kramer2007} (see~\cite{abs-1908-11179}). 

\subsubsection{Specifying and Verifying New Adaptation Goals and Models}  

\begin{figure}[h!]
	\centering
	\begin{subfigure}[b]{0.43\linewidth}
		\includegraphics[width=\linewidth]{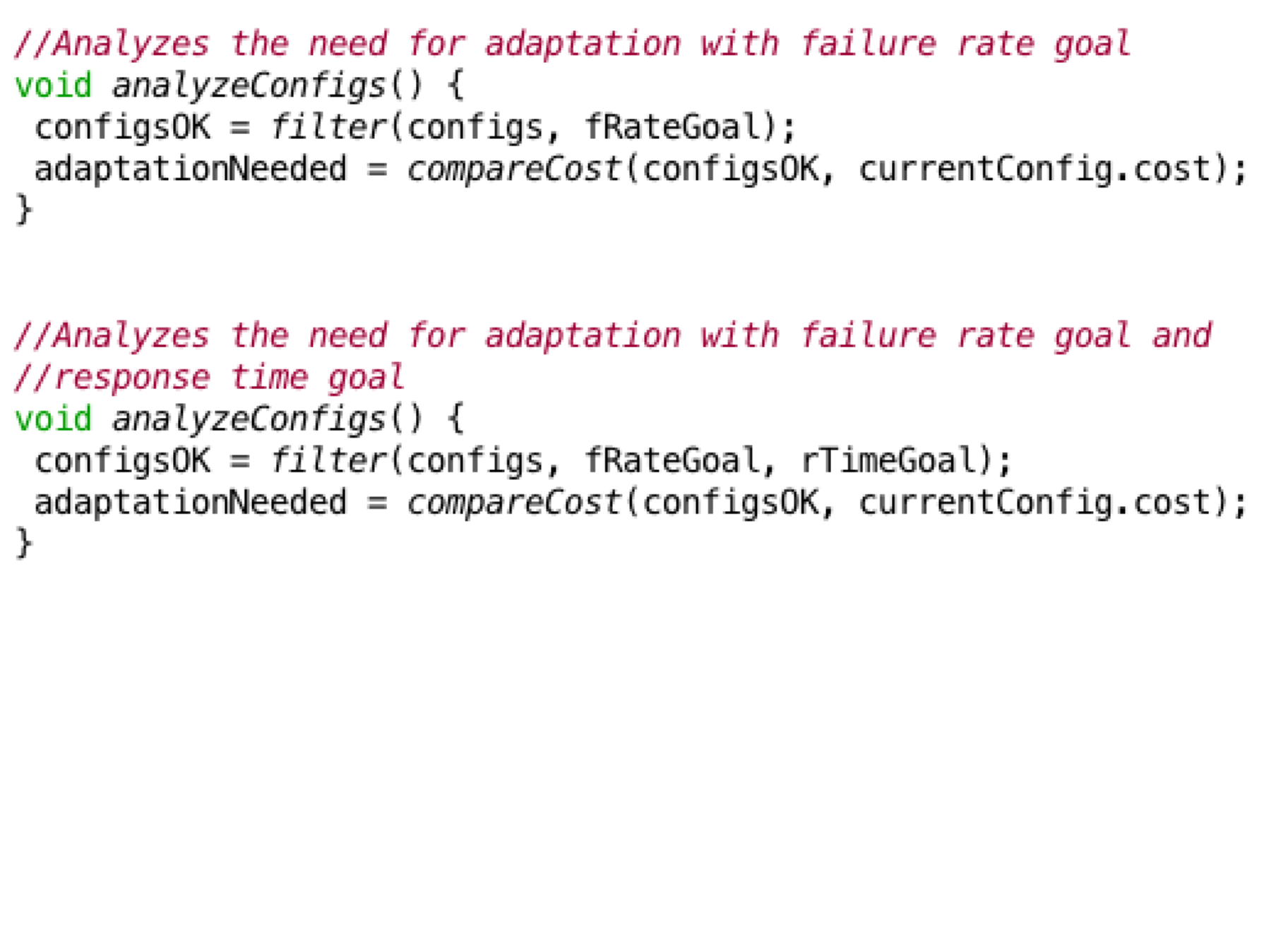}
		\caption{Update of a function in the Analyzer model}
	\end{subfigure}
	\begin{subfigure}[b]{0.56\linewidth}
	\includegraphics[width=\linewidth]{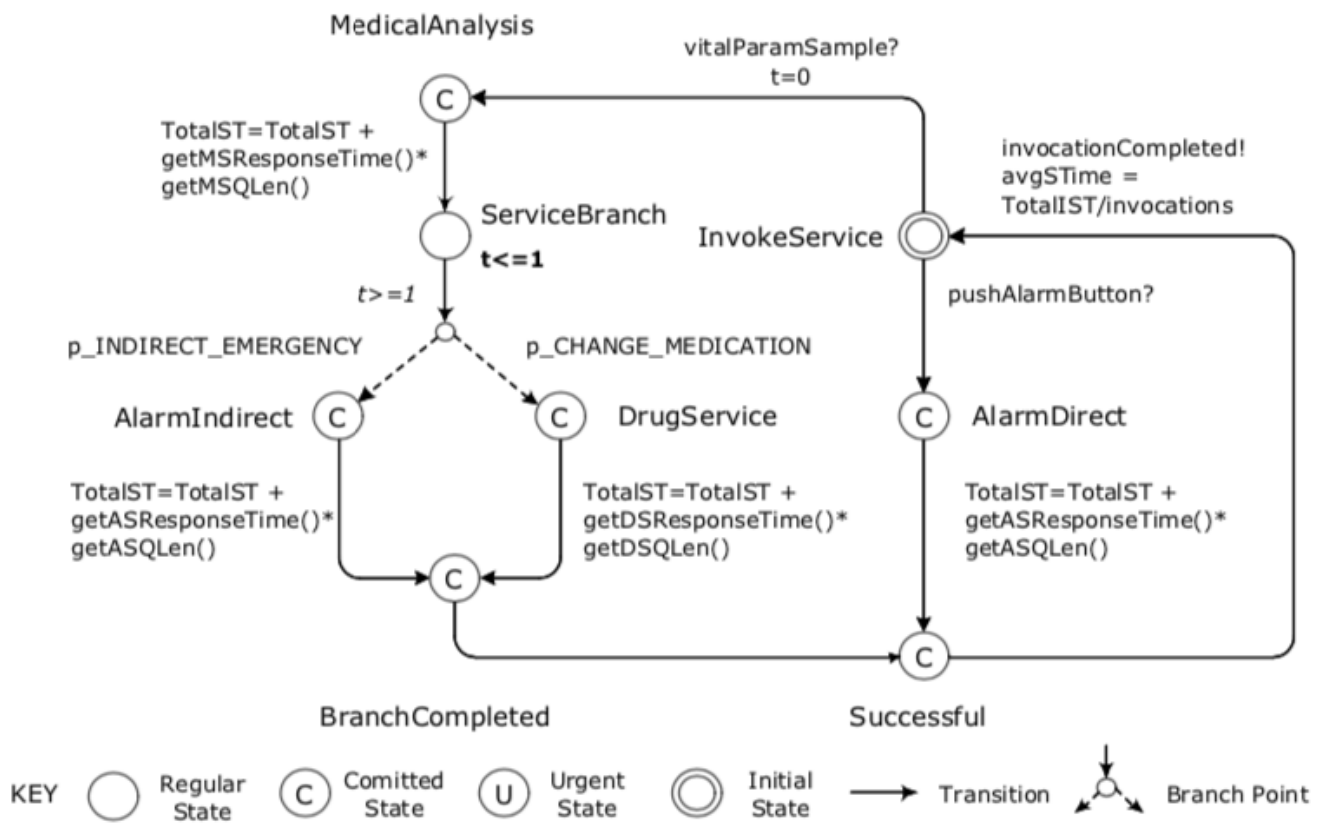}
	\caption{Quality model to predict response time}
    \end{subfigure}
	\caption{Example of an updated and a new runtime model to deal with a new goal in the health assistance system}
	\label{fig:exampleNR}
\end{figure}

We add a new requirement to the health assistance system that keeps the average response time of service invocations under a required value. The corresponding adaptation goal can be specified as a threshold goal, similar to the failure rate goal.  	Fig.\,\ref{fig:exampleNR}(a) shows how one of the functions of the Analyzer model is updated to deal with the new response time goal (\textit{rTimeGoal}). Fig.\,\ref{fig:exampleNR}(b) show the quality model for service response time. This model, that works together with the model of the system and environment (see Fig.\,\ref{fig:exampleMAPEmodels}), is used by the verifier during runtime analysis to predict the expected response time for the different adaptation options.  

\subsubsection{Enact New Models} 

When the evolved feedback loop model is verified it needs to be enacted. Model enactment follows a semi-automatic process that is supported by the goal management layer~\cite{Kramer2007} and the model execution engine. We illustrate an example for the IoT application in the next section.

\section{ActivFORMSi Applied to IoT Application}\label{section:instance}

In this section we provide additional information on two parts of ActivFORMSi, the instantiation of ActivFORMS: (1) the design of the knowledge part of a feedback loop model and (2) the design of stub models. We illustrate these for the IoT system.

\subsection{Design of Knowledge Part of a Feedback Loop}

The MAPE model templates of ActivFORMSi are derived from extensive experience with engineering self-adaptive systems~\cite{Didac2013,Iftikhar2014,Shevtsov2015,Weyns2015c,7573167,Calinescu2017}. 
The model templates are specified with Uppaal~\cite{tutorial04}. 
The knowledge part consists of elements that are shared among the MAPE elements. Listing~\ref{knowledge} shows an excerpt of the ActivFORMSi MAPE model template to specify knowledge. 

\lstset{caption={Definition of Knowledge},label=knowledge}
\footnotesize
{
\begin{lstlisting}
//Knowledge = 
//  {Configuration, Adaptation Goals, Adaptation Options, Plan, Quality Models}
	
//A configuration defines the relevant elements of the managed system, a set of
//quality properties, and the relevant properties of the environment	
<Configuration currentConfiguration>;  
	
//Adaptation Goals
<int PROP;> 
<bool optimizationGoal_I(Configuration gConf, Configuration tConf, int PROP) { }>
//Tests whether a test configuration (tConf) outperforms a given configuration 
//(gConf) regarding a property (PROP)
<int PROP;> 
<bool satisfactionGoal_I(Configuration conf, int PROP) { }>
//Tests whether a configuration (conf) satisfies a given property (PROP) 	
	
//Adaptation Options
type struct { 
 <ManagedSystem option>; 
 <Qualities verificationResults>;
} AdaptationOption
AdaptationOption adaptationOptions[MAX_OPTIONS];
	
//Plan
type struct {
<const stepType>; 
<Element element>; 
<Value newValue>; 
} <Step>
type struct {
<Step steps[MAX_STEPS]>; 
} Plan
	
//Quality Models 
//A network of stochastic timed automata per quality for each of the adaptation goals
\end{lstlisting}
}
\normalsize

$\mathit{Knowledge}$ comprises five elements: the current $\mathit{Configuration}$, a set of $\mathit{Adaptation Goals}$, a set of $\mathit{Adaptation Options}$, i.e., the possible configurations of the managed system, a $\mathit{Plan}$ consisting of adaptation steps that are composed by the \textit{Planner} (the MAPE models are explained below), and a set of $\mathit{Quality Models}$, one model for each quality that is subject of an adaptation goal. 

The adaptation goals define the quality objectives that need to be realized by the feedback loop. ActivFORMSi offers support to model adaptation goals as boolean functions. We distinguish between an $\mathit{optimizationGoal}$ that tests whether a configuration $\mathit{tConf}$ outperforms a given configuration $\mathit{gConf}$ regarding a property ($\textit{PROP}$), and a $\mathit{satisfactionGoal}$ that tests whether a configuration ($\mathit{conf}$) satisfies a given property ($\textit{PROP}$). However, ActivFORMSi is not limited to these types of goals, so other types of goals can be defined and applied. 

An adaptation option consists of two parts: a particular setting of the managed system ($\mathit{option}$) and a placeholder for the verification results ($\mathit{verificationResults}$). The \textit{Analyzer} determines the adaptation options based on the range of settings of elements of the managed system that can be adapted (see Section \ref{subsec:analysis}). The verification results are added when the verifier has produced estimated values for the different qualities per adaptation option. The \textit{Planner} then picks the best option based on the verification results using the adaptation goals. In this paper, we assume that a limited but possibly large number of adaptation options are available when adaptation is required ($\textit{MAX\_OPTIONS}$). This implies that any system parameter that can be used for adapting the managed system with a value in a continuous domain needs to be discretized and limited in range. Heuristics can be applied to select the adaptation option from a very large set, but this is out of scope of this paper.  

A plan consists of a series of steps ($\mathit{Step}$), each defined by a $\mathit{stepType}$, an $\mathit{element}$ (optionally refined by sub-elements) that refers to an element (a parameter, algorithm, component, etc.) of the managed system to which the step applies, and the $\mathit{newValue}$ (a setting, rate, status, etc.) that needs to be applied to the element. 

\begin{figure}[h!tb]
	\centering
	\includegraphics[width=0.8\textwidth]{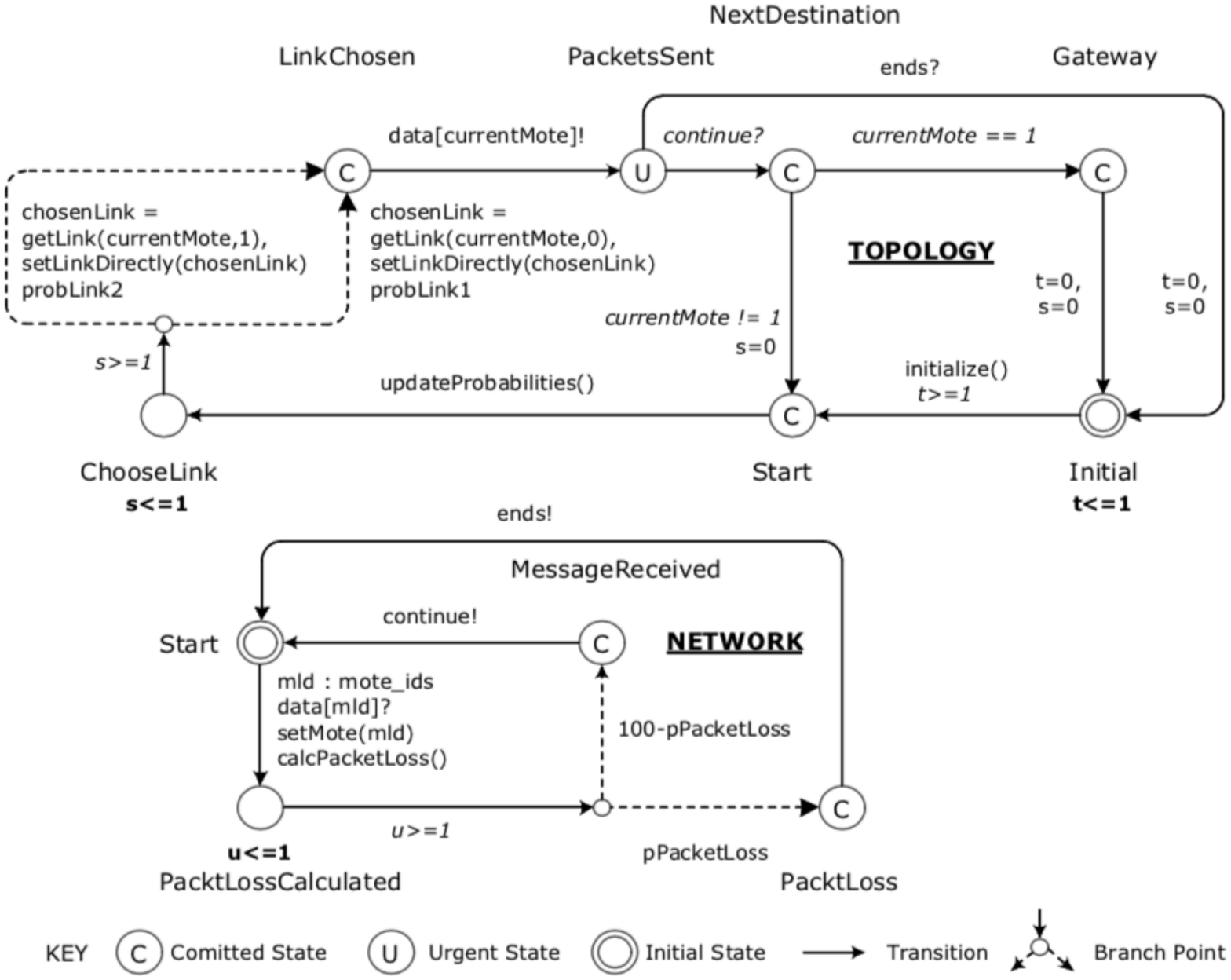}
	\caption{Runtime quality model for packet loss}
	\label{fig:runtime_quality_model_packet_loss}
\end{figure}

The quality models are in essence domain-specific abstractions of the behavior of the managed system and its environment, each model capturing the characteristics of one quality that corresponds to an adaptation goal. The analyzer, supported by the verifier, uses the quality models to perform “what-if analyses,” that is, it determines what would be the expected quality values of the system if a particular adaptation option is selected to adapt the system. To that end, quality models have parameters to set the possible configurations of the managed system, i.e, the adaptation options. Based on the verification results (the estimated quality values per adaptation option) and the adaptation goals, the planner then selects the best option to adapt the system. In ActivFORMSi, each quality model is specified as a parameterized stochastic timed automaton (or a network of these).
\vspace{5pt}\\
\noindent
\textbf{Example.} We illustrate a quality model for packet loss of the IoT system shown in Fig.~\ref{fig:runtime_quality_model_packet_loss}. 

The adaptation options are determined by two parameters: the power settings of all the links of the network and the settings for the distributions of messages of the motes with two parents (i.e., the distribution factors). The power setting (0 to 15) for each link is set based on the actual signal to noise ratio (SNR) along that link. To that end, the model uses link-specific functions:

\begin{quote}
	$SNR_l = \alpha_l + \beta_l \times power_l$
\end{quote}

with $SNR_l$ the actual SNR along link $l$, $power_l$ the power setting of the child mote of link $l$, and $\alpha_l$ and $\beta_l$ two factors that determine the relationship between the power settings and the SNR for that link ($\alpha_l$ and $\beta_l$) are determined based on experimental data derived from observations in the field). The power setting is then set to the minimum value that is required to ensure that the SNR is at least zero, ensuring a low packet loss.
The distribution factors for links of motes with two parents are set from 0 to 100\% in steps of 20\% (0,100), (20,80) ... (100,0). These values are assigned to the variables of the topology model ($probLink1$ and $probLink2$).
The messages are then distributed probabilistically based on the values assigned for the adaptation option that is verified. For motes with one parent, the probability of one link ($probLink1$) is set to 100, so this link is selected. 

Furthermore, the values for two types of uncertainties need to be set: the traffic load generated by the motes and the signal to noise ratio per link (SNR). These uncertainties values apply to all adaptation options at a given point in time. A number of motes generate a steady traffic load (i.e., motes 3, 8, 9, and 15 that periodically sample the temperature, see Fig.~\ref{fig:DeltaIoT}). The load generated by these motes is represented by constants. Other motes generate a fluctuating traffic load (i.e., based on the presence of humans, e.g., motes 4, 8, and 10, see Fig.~\ref{fig:DeltaIoT}). The loads of these motes are determined probabilistically based on profiles derived from field experiments. Similarly, the values of the SNR per link that depend on external factors such as network interference and noise in the environment are determined based on profiles. These values are used to determine the transmission power settings of the motes per link as explained above. The values for traffic load and SNR are periodically collected by the gateway and updated by the probe.

We now briefly explain the models. When the \textit{Topology} automaton is triggered, data is sent along a sequence of links to the \textit{Gateway} (see also Fig.~\ref{fig:DeltaIoT}). When communication starts (\textit{initialize()}), a link is selected (\textit{ChooseLink}) determined by the distribution factors. When a link is selected, the model signals the \textit{Network} automaton (\textit{data[currentMote]!}). The network automaton receives the signal (\textit{data[mId]?}) with the identifier of the mote that sends data (\textit{mId}). The probability for packet loss is then calculated using the \textit{calcPacketLoss()} function. The probability that packets get lost during communication depends on the SNR for the link (as explained above). Depending on the value of the packet loss either the transition \textit{PacketLossCalculated} to \textit{PacketLoss} is taken (communication failed) or the transition \textit{PacketLossCalculated} to \textit{MessageReceived} is taken (communication was successful). After a successful communication, the network automaton returns to the \textit{Start} location. The Topology automaton will then continue with the next hop of the communication along the path that is currently checked, until the Gateway is reached (\textit{currentMote == 1}). If a packet gets lost, the communication along the path that is currently checked ends. As such, the quality model allows determining the packet loss of the adaptation options by performing simulations of the communication of packets through the network taking into account the current uncertainties until results with the required accuracy and confidence are obtained.

For the design of the MAPE models, we refer to~\cite{abs-1908-11179} for a specification of the templates and its application to the IoT system.

\subsection{Design of Stub Models}

\begin{figure}[h!tb]
    \centering
    \includegraphics[width=\textwidth]{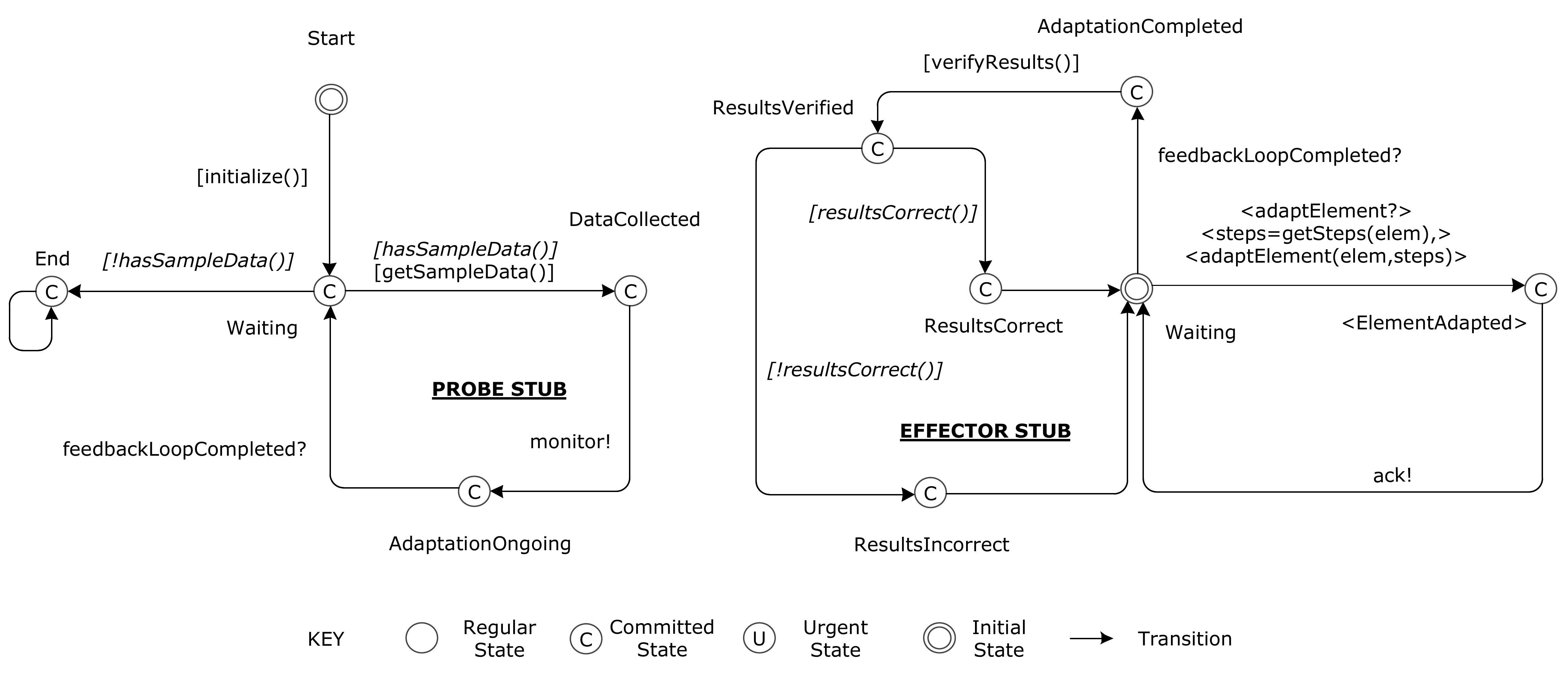}
    \caption{Templates for probe and effector stubs}
    \label{Stub_templates}
\end{figure}

Verifying the MAPE models require stub models. ActivFORMSi support engineers with a set of generic templates to define these stubs for the adaptation problem at hand. Figure~\ref{Stub_templates} shows a stub templates for a probe and effector; others templates are available at the ActivFORMS website. 

We illustrate a few scenarios for verifying the correctness of the MAPE models for the IoT system. The excerpt of the probe stub in Listing\,\ref{stubsDeltaIoT} shows how the designer specifies the initial configuration together with the initial quality properties of the network. Then follows the sample data that applies a series of changes to the configuration and network properties; in sample 2 for example, the value of the packet loss is increased with 20\%. The verifier stub  determines for each run the values for the quality properties of all the adaptation options. Each sample type covers a specific trajectory in the MAPE models. For instance, when the packet loss in the network is increased with 20\% the quality estimates produced by the verifier will require the planner to find a new best adaptation option and prepare a plan for this. On the other hand, when the SNR of link is reduced with 5 dB, the quality estimates produced by the verifier will not require adaptation (i.e., the current configuration is the best option). The different sample types cover scenarios with complete and partial verification. The excerpt of the effector stub shows how the correctness of the adaptation can be checked, i.e., the estimated qualities of the selected configuration comply with the adaptation goals and the best adaptation option has been applied. 

\lstset{caption={Excerpts of stubs used for the verification of MAPE models for DeltaIoT},label=stubsDeltaIoT}
\footnotesize
{
	\begin{lstlisting}
//EXCERPT PROBE STUB
DeltaIoT sample = {
 //{ID, load, parents, {source, destination, power, SNR, distribution}
 ...
 {7, 10, 2, {{7, 2, 15, 3, 100}, {7, 3, 15, -3, 100}}, 
 ...
}
//{packetLoss, energyConsumption, latency}
QoS qos = {5, 25, 0}; 

void getSampleData(){
 if (sampleCount == 0){
  deltaIoT = sample; } 
 ... 
 else if (sampleCount == 2) {
  qos.packetLoss +=20; }
  ...
 else if (sampleCount == 5) {
   deltaIoT.motes[3].links[0].SNR -=5; }
 ...
}

//EXCERPT VERIFIER STUB
void getSampleQualityEstimates(){
 ...
 //Full verfication; some adaptation options satisfy goals
 if (sampleType == 1){
  for(i=0; i < Knowledge.adaptationOptions.size; i++) {
   adaptationOptions.options[i].verifResults.packetLoss = pLoss(sampleType, i); 
   adaptationOptions.options[i].verifResults.energyConsumpt = eCons(sampleType, i);
  }
 }
 ...
 //Partial verification, some adaptation options satisfy goals 
 else if (sampleType == 4){
  for(i=0; i < Knowledge.adaptationOptions.size; i++) {
   adaptationOptions.options[i].verifResults.packetLoss = pLoss(sampleType, i); 
   adaptationOptions.options[i].verifResults.energyConsumpt = eCons(sampleType, i);
  }
 }
 ...
}

//EXCERPT EFFECTOR STUB
bool result; 
void verifyResults(){
 ...
 result = estimatedQualitiesEnsureAdaptationGoals(Knowledge.bestAdaptationOption) &&
 bestOptionApplied(Knowledge.currentConfiguration, Knowledge.bestAdaptationOption);   
}
bool resultsCorrect(){
 return result;
}
	\end{lstlisting}
}
\normalsize

\section{Evaluation of ActivFORMSi}\label{section:evaluation}

We evaluated ActivFORMS and its tool-supported instance using the DeltaIoT network deployed at KU Leuven, shown in Fig.~\ref{fig:DeltaIoT}. We zoom here in on one particular aspect: a tradeoff analysis between accuracy and adaptation time. For the detailed description of the evaluation setting and other evaluation results we refer to~\cite{abs-1908-11179} and the ActivFORMS website. 
\vspace{5pt}\\
\noindent 
\textbf{Tradeoff Between Accuracy and Adaptation Time.}\label{tradeoff}
To evaluate the tradeoff between the accuracy of the verification results and adaptation time (which is primarily determined by the verification time), we used a network with 15 motes and two adaptation goals: energy consumption and packet loss. First, we evaluated the tradeoff between accuracy and confidence of the verification results and verification time. Second, we evaluated the quality of adaptation decisions for different settings.

\paragraph{Results}

\begin{figure}[h!]
	\centering
	\includegraphics[width=\textwidth]{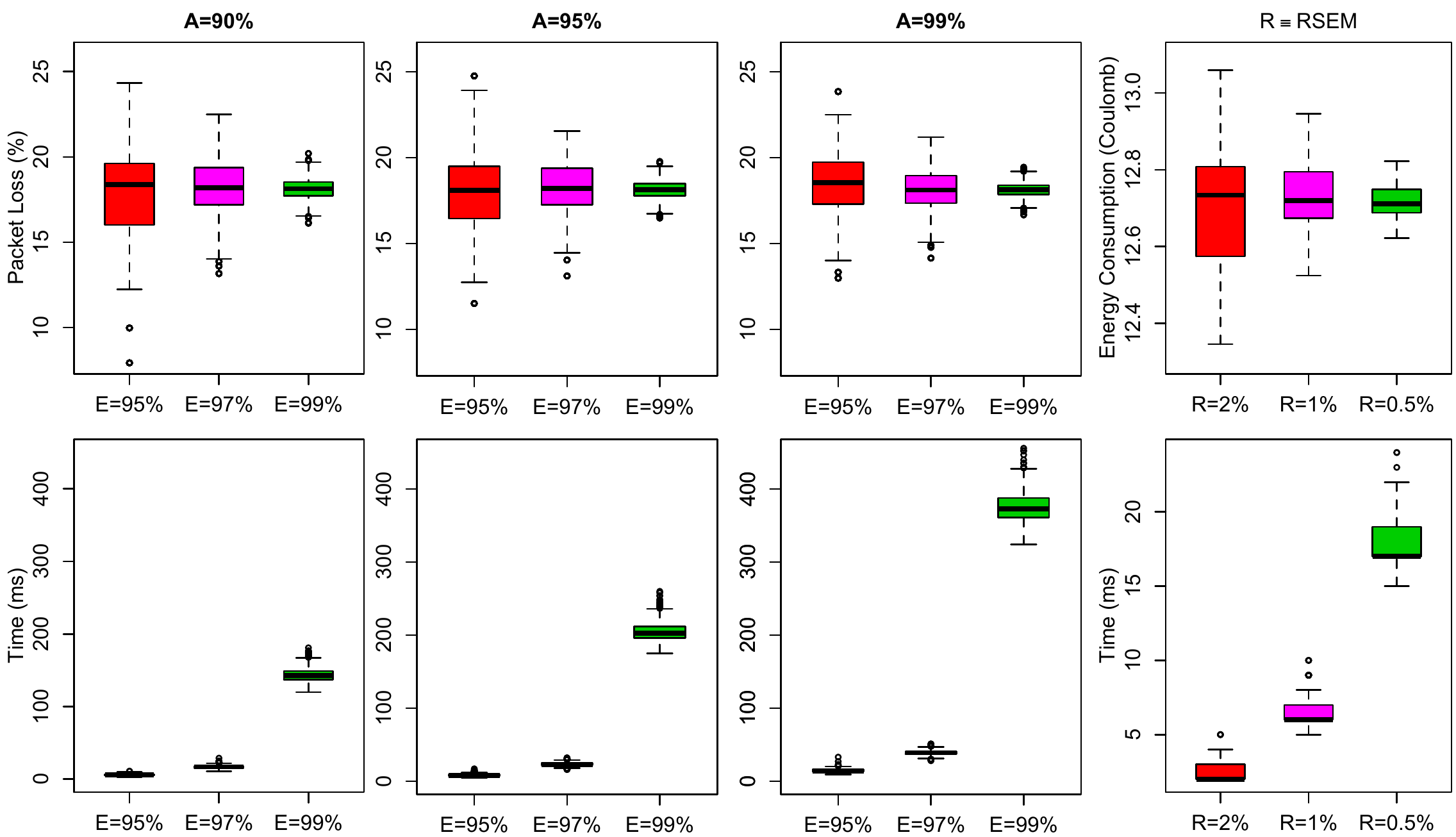}
	\caption{Tradeoff between accuracy and adaptation time with ActivFORMS}
	\label{fig:tradeoff-1}
\end{figure}

In the first experiment, we picked a random adaptation option and applied verification for both qualities.\footnote{We repeated the experiment for a randomly selected sample of 10\% of the adaptation options with randomly assigned values for the uncertainties. These experiments provided similar results. A report with the results is available at the ActivFORMS website.} The graphs in Fig.\,\ref{fig:tradeoff-1} plot the results of 10K runs. The results for packet loss (graphs on the left hand side) show the effect of accuracy $E$ (that defines the approximation interval [$p\pm\epsilon$]) and confidence $E$ (defining $1$\,-\,$\alpha$) of the verification results. The results show that higher accuracy and confidence settings provide better verification results. E.g., the quartiles of the box plot for a setting with $95\%$ confidence and $95\%$ accuracy are $1.64$/+$1.41\%$ compared to $-0.35$/+$0.36\%$ for a setting with the same confidence and $99\%$ accuracy. If we increase the confidence from $95\%$ to $99\%$ with the same accuracy of $95\%$ the quartiles of the boxplot are $-1.24$/+$1.21\%$. 
The tradeoff is an increase in verification time. In particular, increasing the accuracy from $95\%$ to $99\%$ with the same confidence of $95\%$ increases the verification time from $8.37$ms to $204.22$ms (mean values). On the other hand, 
increasing the confidence from $95\%$ to $99\%$ with the same accuracy of $95\%$ increases the verification time from $8.37$ms to $14.50$ms (mean values). These results show that increasing the accuracy has a larger effect on the quality of the verification results compared to increasing the confidence. However, it has also requires more verification time. These results confirm the impact of accuracy and confidence on the verification time for SMC~\cite{David2015}. 

The results for energy consumption (graphs on the right) are similar; more accurate verification results (i.e., lower values for RSEM) require more verification time (i.e., more simulation runs). E.g., the quartiles of the boxplots for $RSEM$\,=\,$2\%$ are -$0.16$/+$0.07$~C compared to -$0.02$/+$0.04$~C for $\mathit{RSEM}$\,=\,$0.5\%$. The cost is an increase of average verification time from $2.51\,ms$ for $\mathit{RSEM}$\,=\,$2\%$ to  $17.83\,ms$ for $\mathit{RSEM}$\,=\,$0.5\%$. The results for other adaptation options are similar, we refer the interested reader to the ActivFORMS website for the results. 

\begin{figure}[h!]
	\centering
	\includegraphics[width=.8\textwidth]{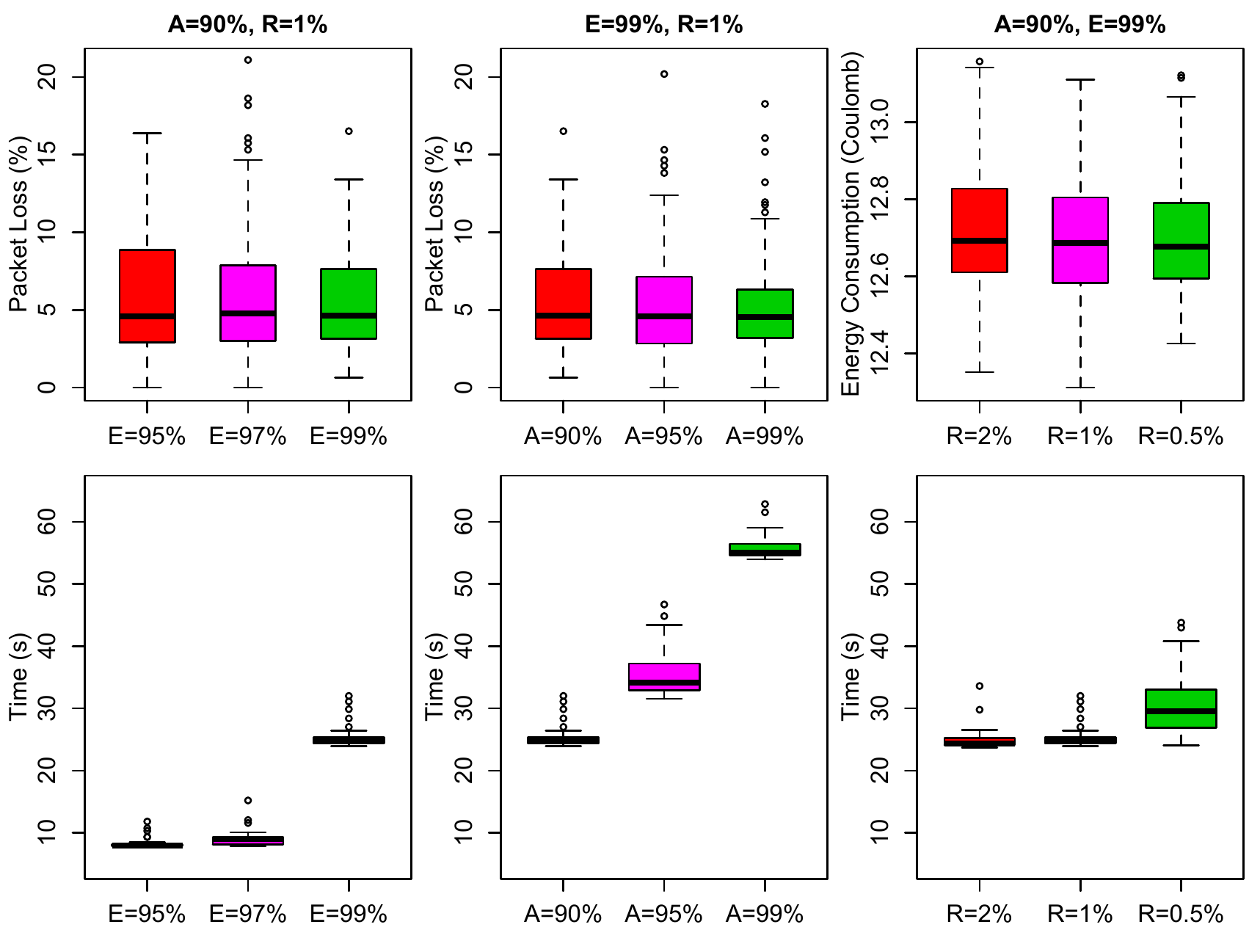}
	\caption{Impact of verification settings on quality properties (E: accuracy; A: confidence; R: RSEM)}
	\label{fig:tradeoff}
\end{figure}

In the second experiment, we evaluated the quality of adaptation decisions and verification time for a simulation run of $12$ hours for the different settings of accuracy $E$, confidence $A$, and RSEM. Fig.\,\ref{fig:tradeoff} shows the results. The boxplots show that settings with higher accuracy, confidence, and RSEM produce more accurate verification results and hence better adaptation decisions. For example, for a setting with $A$\,=\,$90\%$ and $E$\,=\,$95\%$ ($\mathit{RSEM}$\,=\,$1\%$), the quartiles for packet loss are -$1.69$/+$4.27\%$, compared to -$1.34$/+$1.77\%$ with both $E$ and $A$ set to $99\%$. The cost is an increase of adaptation time from $8\,s$ to $55\,s$ (mean values). For energy consumption, the quartiles for a setting with $\mathit{RSEM}$\,=\,$2\%$ ($A$\,=\,$90\%$, $E$\,=\,$99\%$) are -$0.08$/+$0.14$~C, compared to -$0.08$/+$0.11$~C for a setting with $\mathit{RSEM}$\,=\,$0.5\%$. The cost is an increase of adaptation time from $24\,s$ to $30\,s$ (mean values).

\paragraph{Conclusions} 
We can conclude that applying runtime statistical model checking to the default DeltaIoT setting with settings that produce smaller approximation intervals and higher confidence result in better adaptation decisions, but the cost is an increase of adaptation time. The effect of the approximation intervals on the verification time is higher as the effect of confidence, confirming the basic principles of SMC. Dealing with this tradeoff is a domain-specific problem and depends on the requirements at hand.

\section{Opportunities for Future Work}\label{section:conclusions}

Guaranteeing that a self-adaptive system behaves correctly and ensures the adaptation goals in an efficient way is challenging. To that end, we presented ActivFORMS (Active FORmal Models for Self-adaptation), a end-to-end approach for engineering self-adaptive systems with guarantees. 
Opportunities for future work include: 
\begin{itemize}
 \item Study the use of online learning techniques to deal with large adaptation space, i.e., runtime configurations with a largen number of adaptation options. A key challenge will be to define the impact on the guarantees that can be obtained. Starting points are\,\cite{QuinWBBM19,VanDerDonckt2020,GheibiWQ21}. 
 \item Study how ActivFORMS can be applied to adaptation problems with more complex types of uncertainties, such as uncertainties of the structure of models and uncertainties in cyber-physical settings\,\cite{BuresWSTBGGGKKP17,MusilMWBMS17,8787077,WeynsBCCFGNPRRS21}. 
 \item Study how ActivFORMS can be applied in systems with multiple feedback loops that need to work together to solve an adaptation problem\,\cite{WeynsSGMMPWAGG10}. The authors of \cite{QuinWG21} provide an overview of the state of the art in decentralized self-adaptive systems. Inspiration can also be taken from the field of multi-agent systems and coordination, see for instance\,\cite{Wooldrige2009,weynsmichel}, and patterns for multi-agent systems, see for instance\,\cite{Schelfthout02agentimplementation,Sabatucci06,JuziukJoanna2014DPfM}.

 \end{itemize}


\bibliographystyle{ACM-Reference-Format-Journals}
\bibliography{referenser}

\end{document}